# Project Severe Weather Archive of the Philippines (SWAP) Part 1: Establishing a Baseline Climatology for Severe Weather across the Philippine Archipelago

Generich H. Capuli[*,1]

[1] Department of Earth and Space Sciences, Rizal Technological University, Brgy. Malamig, Mandaluyong City, Metro Manila 1550, Philippines

## Abstract

Because of the rudimentary reporting methods and general lack of documentation, the creation of a severe weather database within the Philippines has been difficult yet relevant target for climatology purposes and historical interest. Previous online severe weather documentation i.e. of tornadoes, waterspouts, and hail events, has also often been few, inconsistent, inactive, or is now completely decommissioned. Several countries or continents support severe weather information through either government-sponsored or independent organizations. For this work, Project SWAP stands as a collaborative exercise, with clear data attribution and open avenues for augmentation, and the creation of a common data model to store the phenomenon's information will assist in maintaining and updating the aforementioned online archive in the Philippines. This paper presents the methods necessary for creating the SWAP database, provide broader climatological analysis of spatio-temporal patterns in severe weather occurrence within the Philippine context, and outline potential use cases for the data. We also highlight the project's current limitations as is to any other existing and far larger database, and emphasize the need for understanding these events' and their mesoscale environments, inline to the current severe weather climatologies across the globe.

## 1. Introduction

Convective storms such as thunderstorms are capable of producing large hail (*yelo*), extreme rainfall (*Labis na pag-ulan*), and tornadoes/waterspouts (*buhawi/ipo-ipo*). It can be destructive to manmade structures and can lead to loss of life, especially in urban areas with high population, building density, and variable quality. This is an important challenge for meteorologist as such forecasts require knowledge of the environment of the storms, which can be obtained from radiosonde measurements or numerical weather prediction models. However, tallying the impacts of each severe weather events (SWEs) that occurred poses yet another challenge due to the problems associated with non-meteorological effects in data collection (Brooks et al., 2014) which is why the approach to data acquisition needs to be standardized to reduce potential error sources (Concannon et al., 2000; Antonescu et al., 2017).

Archives of SWEs, mostly on tornadoes, occurrence are currently spread among individual countries and regional research groups, with varying quality, standards, and temporal coverage. Thus, it is difficult to draw conclusions about the worldwide spatial or temporal distribution of tornado frequency or strength. Efforts to estimate a global climatology have been made, notably by Fujita (1973) and by Goliger and Milford (1998), but relied on fragmented and sometimes contradictory sources. The more recent European Severe Weather Database (ESWD), published by the European Severe Storms Laboratory (ESSL, Dotzek et al., 2009), has been used to sometimes contradictory sources and has been used to estimate tornado climatologies for Europe (Grieser and Haines, 2020) and now integrated well to the global archives by Maas et al. (2024).

The production of an official, organized database of extreme weather reports is essential to identify the nature of these phenomena. An inventory of events is the first step to identify threats, and it is valuable information for planning and insurance matters (Brooks and Doswell, 2001). For this reason, several countries and regions in the world have generated severe weather and tornado databases, including China (Chen et al., 2018), Italy (Miglietta and Matsangouras, 2018), Portugal (Leitão and Pinto, 2020), the Czech Republic (Brázdil et al., 2019), the British Isles (Mulder and Schultz, 2015), and the USA (Guo et al., 2016), among others. These databases are mainly supported and maintained by their national agencies, such as national meteorological services e.g. Environment and Climate Change Canada (ECCC, 2015), National Centers for Environmental Information (NCEI, 2024), Riesco Martín et al. (2015), and, in some other cases, as part of academic projects e.g. Dias (2011) and Sioutas (2011).

Similarly, extensive climatologies have been conducted on the incidence of waterspouts in Europe and North America (Sioutas et al., 2013). In Catalonia, Spain, the information obtained from social networks was essential to improve the climatology of tornadoes and waterspouts. It led to a better understanding of the frequency and distribution of waterspouts and severe weather (Rodríguez et al., 2021). The relatively high incidence of waterspouts around the Florida Peninsula led to an attempt to predict these events by applying a statistical model that considers a significant number of variables (Devanas and Stefanova, 2018). The regional predictions of the model on the probability of waterspout incidence were better than some of the applied indices. In summary, research on waterspouts has become relevant in the last decade.

Furthermore, due to the significant importance of hail climatology research, in situ measurement at weather stations, rawinsonde measurements, and several kinds of remote sensing data, such as radar and satellite, have been used to analyze the temporal and spatial distributions of hails across different continentals and countries of the world (Allen et al., 2020). There are also several studies about the climatology of hail day (Li et al., 2016), hail frequency and size (Li et al., 2018), large 2" hail (Ni et al., 2020), and the potentially influential factors of hailstorms (Zhao et al., 2018) in China in recent decades collected and provided by the National Meteorological Information Center.

Besides these professional observation datasets, social databases had also been used to demonstrate the characteristics of hail distributions, such as news reports, disaster addresses from insurance companies, records of the agriculture and housing industry, and statistical yearbooks. Tuovinen et al. (2009) investigated the severe hails in Finland by collecting newspaper, storm spotter, and eyewitness reports in 70 years. The ESWD which is a database of SWEs reported from crowdsources in 2002 (Dotzek et al., 2009). The Tornado and Storm Research Organization (TORRO) holds a website to let voluntary persons upload severe weather reports such as tornadoes, lightning and damaging hailstorm on the island of Britain (Webb et al., 2009). However, damage quantification and characterization are more problematic due to the meteorological context in which each case occurs, e.g. events occurring during a severe storm. Such issues need to be considered to establish the limitations of baseline climatology within the country based only on documentary evidence.

In the Philippines, however, no previous climatologies of SWEs have been made, nor have reports been systematically collected previously by the Department of Science and Technology-Philippine Atmospheric, Geophysical, Astronomical Services Administration (DOST-PAGASA), though situational reports and assessments still exists in few cases. This lack of research is probably because of the small number of interested scientists and the general misassumption that severe convective weather rarely occurs. In fact, reports of severe weather in a country, that country's perception of its severe convective storm potential, and its forecast and warning process are intimately linked (Doswell, 2003). The collection of severe weather reports are useful to educate people about the potential risk of severe weather and verification of weather warnings. Thus, the present study provides a unique opportunity to examine the occurrence of severe weather at the tropical latitude.

This paper establishes the first, simple database for SWEs and a baseline climatology that include tornadoes, waterspouts, and hailstorms based on data compiled from hemerographic sources and personal communications, which contributes to the knowledge of climate types in the Philippines through Project Severe Weather Archive of the Philippines (SWAP). Now in its 2nd Data Release (labelled as SWAP DR2), we utilized the compiled data to investigate the distribution of severe weather occurrence and record-keeping, as well as the interdependence between the two. We present the compiled database as-is, acknowledging known biases existing therein. As the true severe weather climatology is fundamentally intertwined with the historical context surrounding its observation, we analyze trends in our database with both in mind.

# 2. Methodology

## 2.1 Data Sources: Tornado, Hail, and Waterspout database

The documentary information was collected from official and non-official sources. The first type of data contains reports from the DOST-PAGASA, National Disaster Risk Reduction and Management Council (NDRRMC), Department of Social Welfare and Development and their Disaster Response Operations Management, Information and Communication sector (DSWD-DROMIC), and local civil protection units distributed throughout the country. These are the official sources that were also included in the archive and are available through their respective platforms, including the National Library of the Philippines (NLP) and Relief Web.

Information not included in any major available dataset was obtained from other reliable sources as indicated in De Coning and Adam (2000), which were used both to add new tornadoes and fill in missing data for existing ones. Doing so is quite time-consuming, as a vast number of different sources are involved, and the effort to find and parse them is continuing. The non-official sources include eyewitness reports, media news/newspapers[1], and information from social media platforms (e.g. Twitter, Facebook, and YouTube). Such documentary data were exhaustively evaluated to identify fake events. The updated database includes, as far as possible, information about the type of event, date, hour, location (latitude-longitude coordinates), tags e.g. tornadoes, hails, elevated, high-based etc., photographs and/or videos through the documentation column, and other relevant information. 46% of the total collated reports (233 tornado reports, 27 hail reports, and 14 waterspout reports) do not contain photographs and/or videos, but was considered in the database as it was thoroughly described from the official and non-official sources highlighted in the 'Notes' column as "*No pictures*".

In addition, some problems occurred with several eyewitness reports (particularly individuals from social media platforms such as Facebook) because of the lack of details, mistaken time, date, or even year of report. In these cases, a request to clarify the information was sent back to the observer. If clarification was not received, the report was rejected. To regulate the process, every SWEs in the database was given a mandatory "Source" attribute which either referred to one of our major datasets or another source (usually a web link). We designated these sources whether they are and still functioning (*Active*), preserved somewhere else like a library (*Preserved*), or no longer accessible or maintained (*Inactive*).

The official record on damage to public and residential infrastructure is limited to only a few tornadoes e.g. the Manila City and Bacolor, Pampanga Tornado events. Despite the scarcity of information about the damage caused by each tornado or waterspout in the database, a classification using, for instance, the Enhanced Fujita Scale (EF) was carefully considered, but should be updated upon further review. Although there are inconsistencies to the details provided by eyewitnesses and media news, in addition to the lack of instrumentation for immediate forecasting, the structural differences in the buildings in Philippines compared to those considered in the EF scale, the limited capacity to conduct field investigations, and the nonexistence of a meteorological observer network, we intend to provide initial ratings to these events by examining texts and the damages it caused through photographic/videographic evidences.

## 2.2 Project SWAP's Data Integration Process

Building the Project SWAP and its dataset of severe weather in a populated country, though had no formal mechanism for reporting severe weather for many years is challenging and time consuming. In addition, creating a climatology requires multiple approaches to get enough data for satisfactory results. The integration of these diverse sources; both official and non-official sources, is essential for creating a holistic view of SWEs across the archipelago and to growing the dataset in the future. Figure 1 displays the data integration process that Project SWAP implemented in archiving SWEs across the Philippines.

---

1 Most of these newspapers are available at the National Library of the Philippines upon request. Same goes to other official sources such as those from DOST-PAGASA whose documents are very difficult to find in the internet.

Once a SWE occured, posted, and/or reported, initial binning scheme takes place classifying the report either as an official or non-official source, or can be reported through Project SWAP's Contact Form. For most part, non-official sources such as media news and eyewitness reports spread across social media are the major contributors of the project's archive. For Project SWAP, the most important aspect with regards to archiving these events are the location, time-date, and photographic/videographic evidences. In this initial step, the important informations such as location, time-date of occurance, and supporting evidences should be noted and tagged. Additional details were also included in this phase such as the estimated EF rating as discussed before which we provide and for hail reports, its estimated diameter (Project SWAP's Hail Scale is to be discussed in next section). For official sources from the likes of DOST-PAGASA, NDRRMC, and DSWD-DROMIC, including other government units in the local scale, its underlying information is usually precise and well-documented. For unofficial sources however, such as social media posts, the location may need to be inferred or estimated based on user-provided details or geotagging.

Secondly, since official data generally has high reliability, it undergoes a minimal validation process to ensure quality and completeness. However, for non-official sources, while potentially valuable, requires a rigorous authenticity check to prevent inaccuracies. These are cross-verified against timestamped media reports or other reliable entries, whenever possible. Specifically, time and location verification would occur as part of the initial authenticity check. This could include cross-referencing with nearby weather reports, timestamps, or even satellite images available through Jaxa's HIMAWARI Monitor P-Tree System. As indicated above in Section 2.1, in some eyewitness reports, we have to request for a clarification of the underlying information that indicated and showcased the SWE. If the non-official data cannot be verified, lacks supporting evidence, or appears dubious, this is then filtered out at this stage to maintain the archive's credibility. However, we considered and archived some cases were only one detail (e.g. hour, town proper) is missing tagging them as *Inactive* and requires a completely delving to its supporting information for an upcoming update. At the time of this writing, Project SWAP currently intakes a maximum of 3 sources, however in a future rendition and data releases, the number of sources will be increased to more than 5 sources.

Third, each data entry is assigned metadata tags that capture essential details such as web links to the sources, locations, latitude-longitude coordinate pairs, and other metadata. The observed time and dates are converted into YYYY-MM-DD and Coordinated Universal Time (UTC) mainly used across meteorological applications and in accordance to World Meteorological Organization (WMO)'s mandate. Names of sources were also preserved as is such as the name of the official sources and non-official sources e.g. *DOST-PAGASA, DSWD-DROMIC, NDRRMC, ABS-CBN, GMA Network, Cebu Star, Inquirer, Philippine Star* etc. For newspapers, they were enclosed with the issued date in YYYY-MM-DD format.

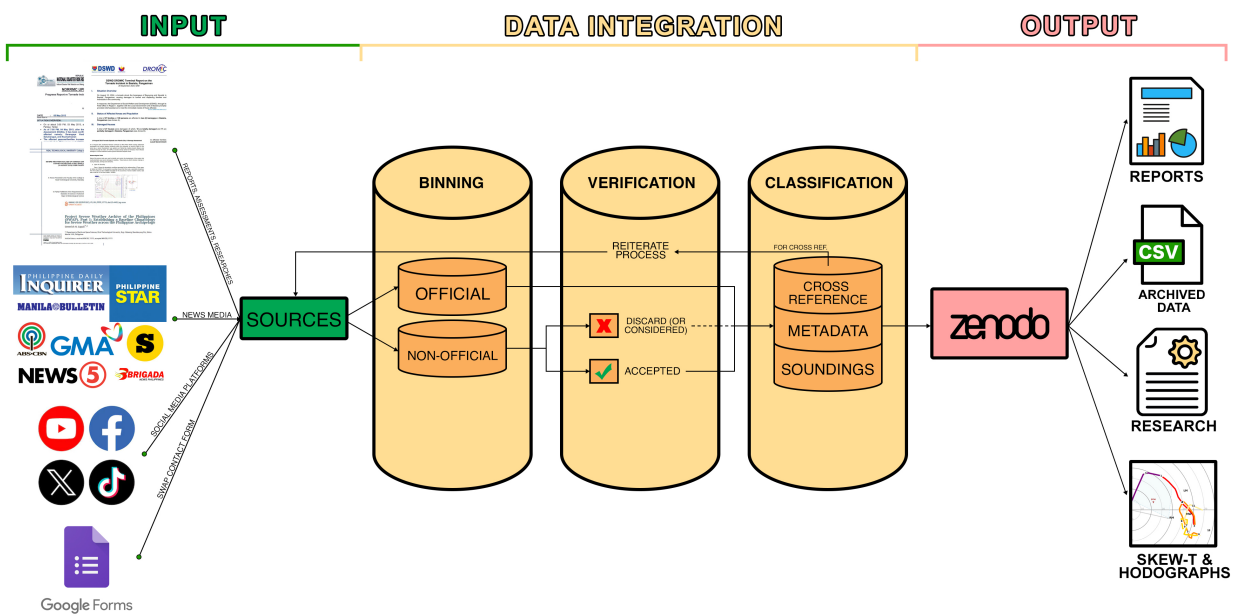

**Figure 1.** The Archiving and Data Integration Process of Project SWAP.

On the other hand, we follow the rules of Philippine Data Privacy Act of 2012 (R.A. 10173) by keeping the name of the eyewitness report anonymous in the documentation columns, but still accessible in relation to the posted SWE information (labelled based of the social media platform they posted). Both official and non-official records that has passed the initial validation stage can also be cross-referenced with other sources to further ensure accuracy. For example, a social media report of a tornado sighting might be cross-checked with nearby weather station records, eyewitness reports, and/or media reports. Once verified and classified, data from both official and non-official sources is integrated into the archive. As also mentioned before, these were also tagged whether they are *Active*, *Preserved*, or *Inactive* as a status indicator for the documentations included.

Project SWAP and its DR2 dataset (including its initial/previous and succeeding releases-outputs) was and will be distributed through Zenodo as we and the aforementioned repository adheres and supports the open science framework. Launched by CERN back in 2013 and funded by the European Commision, Zenodo is popular open and save repository for research data and software that assigns a digital object identifier (DOI) to each release, thus making the Project SWAP citable once being used for another scholarly work.

In general, Project SWAP serves multiple datasets across the public; 3 files are in standard UTF-8 Comma-Separated Values (*csv, comma delimited*) format segmented for the archived tornado, hail, and waterspout reports that can easily be accessed either through Excel for basic checking and both *Pandas* and *GeoPandas* python package as it contains time and date, encompassed region, estimated ratings for tornadoes and waterspouts, estimated hail sizes, location, and latitude-longitude coordinate pairs suitable for in-depth spatial analysis. Another 3 files in Portable Document Format (PDF) also for the archived tornado, hail, and waterspout events each with the inclusion of data charts and hyperlinks to the sources ensuring that each record is traceable to its origin, providing users with a clear understanding of the data's source and reliability level. And finally, a compressed collections of proximity sounding profiles associated to each event to be discussed in succeeding section.

## 2.3 Spatial Analysis and Smoothing

Although Project SWAP, as it currently stands, includes more than half a thousand SWEs, our database is necessarily incomplete, and likely unrepresentative despite covering from 1969-Present due to the lack of other tallied SWEs within the project. A spatial smoother was employed to account for shortcomings in our samplings, and to provide a preliminary climatological estimate of severe event. A non-parametric multivariate kernel density estimation (KDE) was employed to identify and examine hazard patterns/hotspots on a 11 km × 11 km horizontal grid using a kernel function (Silverman, 1986). The general equation for a multivariate KDE with ≥ 2-dimensions given multivariate data set is expressed in Eq. (1)as;

$$\hat{f}(x) = \frac{1}{nh^d} \sum_{i=1}^{n} K\left\{\frac{1}{h}(x - x_i)\right\} \tag{1}$$

Where *K* is the kernel function, in this study, a symmetrical Gaussian kernel was chosen shown by Eq. (2);

$$K(x) = \frac{1}{\sqrt{2\pi}} e^{-t^2/2} \tag{2}$$

The method is particularly useful for analyzing and displaying point data. The occurrences of events are shown as a map of scattered discrete points, which may be difficult to interpret (seen in a later figure). But with the kernel estimation, it generates a density of the events as a continuous field, and thus highlights the spatial pattern as peaks and valleys. The method may also be used for spatial interpolation. In addition, we considered a symmetric mathematical function such as the Gaussian (Normal) metric as it allows for a smooth and continuous representation of density across the spatial field. Closer points i.e. severe weather events contribute more to the density at any location than those farther away capturing spatial trends without sharp transitions. The use of Gaussian kernel was also seen in understanding hail climatology in the U.S. Great Plains (Allen et al., 2019).

Meanwhile, a cross validation was utilized to identify the most optimal bandwidth $(h)^2$, the other variable in Eq. (1), by exhaustively considering parameter conditions, assuming with fixed metric and kernel function i.e. Euclidean points and Gaussian. This is conducted through $k$-fold cross validation with 100 iterations for each 10000 candidates to each of our array of data. Though, a very large grid can be time-consuming to search thorough. Typically, the GridSearchCV (GSCV) offers such capability having 3 compulsory parameters; the estimator, grid of parameters, and number of cross validation $(k)$, and is widely used on machine learning for hyperparameter optimization. However, another way to consider on identifying $h$ is to utilize an adaptive bandwidth with different $h$ values based on the local density of the observations under consideration (Breiman et al., 1977; Abramson, 1982).

# 3. Data Discussion

## 3.1 Limitation of Project SWAP

### 3.1.1 Number of SWEs

Many SWEs go unreported or unrecorded, especially in areas or time periods with less infrastructure or lower population density (Aguirre et al., 1994; Anderson et al., 2007). This is reflected, for example, prior to 2010s including the years 2010 and 2016, were only 2 and 8 cases were tallied to the archive, respectively. We are not sure why it is difficult to look for SWEs way back on Year 2010, but the lack of cases in Year 2016 was likely due to the 14 August 2016 Manila Tornado that occurred on the Metro, making it a standout SWE signature that time, while the rest of the SWEs were concealed by the alias.

The more general increase throughout the time period is due to a combination of better access to more recent tornado records and population expansion that allowed more tornadoes to be observed. Similar patterns are present in most other datasets e.g. ESWD (Dotzek et al., 2009), but they are unlikely to be physical trends, and instead reflect societal growth over time. Under reporting of the total tornado number can also occur when tornado families are judged to be individual tornadoes (Doswell and Burgess, 1988).

In historical reconstructions of tornado, hail, and waterspout events, where the best available sources are newspaper reports written by journalists rather than meteorologists, severe wind events such as microbursts can be misidentified as tornadoes. Reports of tornadic activity are also sometimes sensationalized (Bradford, 1999). Still, the under-reporting effects previously discussed are generally of much larger magnitude, meaning that any observed tornado count for a given region, especially in the pre-modern era, is very likely an underestimate.

### 3.1.2 Tornado and Waterspout Intensity, and Hailstone Sizes

Most of the waterspout events within Project SWAP have occurred on open waters without any damage to property. Thus, was straightforwardly rated as EF0. A tiny fraction of the waterspout events were rated as EF1 which impacted several homes made of light construction materials and shanties.

The difficult part is initially rating more than 300 tornadoes included in Project SWAP. Around a half of total tornadoes were initially rated by this project as EF1, given that most of the infrastructures around the country fall under *One- or two-family residences*, *Low-rise (1-4 story) bldg.*, and both *Hard* and *Softwood Trees* damage indicators. A more comprehensive survey/assessment to update these initial ratings is needed to rectify the initial ratings given.

Following the TA's guidelines Maas et al. (2024), tornadoes for which an intensity rating is unassigned are given EFU (EF-Unknown), which the other half of the tornado reports in this project were recorded. This occurs when a tornado is con firmed (e.g., visually), but no damage is found (NOAA, 2021). In other cases, tornadoes may be rated EFU because there were no damage assessment conducted by professional surveyors, and in some databases e.g. León-Cruz et al. (2022) and National Institute of Water and Atmospheric Research (NIWA, 2021), the tornadoes in

---

2 Also known as the smoothing parameter. The resultant probability density function (PDF) is sensitive to the chosen bandwidth. One can also either use Scott's and Silvermann's estimation methods.

| EF scale | No of Tornadoes | No. of Waterspouts | Hail scale | No. of Hail Events |
|----------|-----------------|--------------------|------------|--------------------|
| EFU | 152 | 0 | Undefined | 27 |
| EF0 | 12 | 132 | Small ≥ 1 cm | 74 |
| EF1 | 156 | 6 | Severe ≥ 2 cm | 18 |
| EF2 | 11* | 0 | Large ≥ 3 cm | 13 |
| EF3 | 0 | 0 | Sig. Hail ≥ 5 cm | 3 |
| EF4 | 0 | 0 | | |
| EF5 | 0 | 0 | | |

*Note: EF2 cases are subject to further and extensive re-evaluation, thus can be downgraded in the archive. However, we consider that some of these events are applicable to have such initial rating (e.g. 1991 Lantapan, Bukidnon Tornado that led to the signing of Proclamation No. 769).

**Table 1.** Distribution of documented SWEs based on the EF scale and hail diameter scale adopted.

this category are the majority. Furthermore, the National Centers for Environmental Information/Storm Prediction Center (NCEI/SPC, NCEI, 2024) until 2016 rate tornadoes that did not cause damage as EF0.

Other avenues to provide damage estimates for tornadoes are either by adopting the International Fujita (IF) Scale developed by the ESSL (Groenemeijer et al., 2023) or developing a damage scale based on the structures suited in the Philippines. Recently, the use of the IF scale (preliminary ver.) was demonstrated and evaluated by Púčik et al. (2024) by applying it to a violent tornado case in Czechia on 24 June 2021. The feedback from the surveyors has been used to improve the IF scale providing a more coherent, global framework for rating tornado and convective wind damage. Either this will be adopted by our country's national weather bureau or develop a new system depending on factors such as the unique structural characteristics found in the Philippines, the availability of data for calibration, and the practical applicability of the scale in local contexts is another subject area that is required to touch and ponder on.

Hail sizes were also estimated based on photographic and videographic evidences presented and included for this project. Most authors have indicated that stones of severe sizes have diameters greater than 2 cm (Cintineo et al., 2012; Li et al., 2018) or 2.5 cm (Bowden et al., 2015). While other authors have established the threshold at which hail becomes "large" at 3 cm (Taylor, 1999). A new category called "significant hail" was also introduced for cases in which the severity has been catastrophic for and in which the the diameter has reached at least 5 cm (Tuovinen et al., 2015; Stull, 2017). Therefore, with current support from recent researches and findings, Project SWAP creates a simple hail scale that considers the initial diameter to classify the hail type as;

Small hail: ≥ 1 cm; Severe hail: ≥ 2 cm Large hail: ≥ 3 cm; Significant hail (Sig. Hail): ≥ 5 cm

If a hail event was reported but without any traceable photographic and/or videographic evidences, then we will consider and label it as 'Undefined'. Table 1 depicts the counts of SWEs that include tornadoes, waterspouts, and hail reports with respect to the scaling utilized on this project.

### 3.1.3 Location and Coordinates

Path widths are difficult to measure and rather unreliable at high degrees of precision – for example, exhibiting unnatural, unexplained shifts over time. Pathlength is more easily measured, but in older and even new records can suffer from underestimates, through incomplete surveying of the entire track, or overestimates, through judgment of tornado families as individual tornadoes (Doswell and Burgess, 1988) The only available and thorough tornado track being generated in the country was conducted on the 14 August 2016 EF1 Manila Tornado by stitching both photographic and videographic evidences to create a path length of the tornado along its estimated time it affected a certain area (Capuli, 2024).

Some databases recorded all tornado paths as single point coordinates (e.g. Mexico, Argentina, China), while others included paths of two or more points (e.g. Europe, Canada, Japan). Most of these archives had a mix of both, with the proportions of each depending on the quality of observations. Furthermore, a few datasets, such as those for Argentina and South Africa, record only the town in which a tornado occurred, meaning that the coordinates listed are simply those of the town center.

That being said, we adopted a single point coordinate for this archive, resulting in some level of bias. For documented cases, e.g. Manila Tornado, the coordinates were placed on a well-known infrastructure it impacted such as hospitals, convenience stores, or schools etc. within the area of interest. However, difficult cases were tallied only on the town-level in which a tornado occurred and seen. With no single objective procedure for resolving these biases, we present the database as-is and recommend careful consideration of individual data collection methodologies in future analysis using it. We hope that tornado paths/swaths in most, if not all of the tornado cases, will be included in a future data releases and version of this archive.

On the spatial analysis side, while using the GSCV in selecting the optimal $h$ is a popular technique in using symmetrical kernel functions such as Gaussian KDE as applied (Silverman, 1986; Heidenreich et al., 2013), the GSCV also has drawbacks, particularly the variance of the obtained smoothing parameters calculated for samples drawn from the same distribution is large. It happens that the function has several minimums, often false and far on the side of too small smoothing, sometimes it does not have any minima at all (Hall and Marron, 1991).

Furthermore, the Gaussian KDE presume that the underlying data are naturally continuous, which is frequently not the case. A symmetric kernel may not be fully suitable for discrete bounded datasets and instead, other types of kernels should be used (Vasiliauskas and Beconytė, 2016; Kiessé, 2017). Other types of kernals such as ArcGIS' Quartic function as used by Smith et al. (2012) in constructing climatologies of storm modes can be considered. While GSCV and Gaussian KDE are powerful tools, careful consideration of kernel choice and software limitations is crucial to achieve accurate and reliable density estimates. Still, the results of this study remains valuable offering a critical baseline for further severe weather research and foundational insights into the spatial distribution patterns of SWEs.

### 3.1.4 Sounding Profiles

In the context of a tornado, hail, and waterspout database, these sounding profiles (Skew-T Hodograph) provide critical insights into the environmental conditions leading up to and during SWEs. By analyzing the vertical structure of the atmosphere at various layers, sounding profiles help identify key factors such as instability, wind shear, moisture content, and temperature gradients, which are essential for understanding the potential for severe weather development to occur at such moment.

Inclusion of these profiles in the archive allows researchers and meteorologists to understand atmospheric conditions with the occurrence of tornadoes, hail, and waterspouts, enhancing predictive capabilities and stimulate this research area. These sounding profiles were queried and analyzed through SounderPy of Gillett (2024), either from observational standpoint or reanalysis analogs mainly from ERA5, or in both. Files will have the standard *csv* file format, *sharppy* file importable to the SHARPpy Sounding Software (Blumberg et al., 2017), and a special *cm1* input file formats for Cloud Model 1 simulation models (Bryan and Fritsch, 2002) which will be available alongside SWAP DR2 and future releases.

## 3.2 Temporal Distribution of Severe Weather Activities

In this research, 596 reports were collected as of the SWAP DR2 increasing the documented activity considerably. Thus, the current climatology considers a total of 326 tornadoes, 133 hail events, and 137 waterspouts distributed over 500 severe weather days. SWAP DR2 currently runs from 1969-Present, with period between 1969-1999 were called as 'Pre-2000s' due to the lack of available data (total of 17 tornadoes were documented and indexed).

The annual distribution of severe weather activity in Philippines is shown in Fig. 2. The documented events vary from 5, 15, 25, and 41 events yr$^{-1}$, on average, for the periods 2000-2010, 2000-2020, 2010-2020, and 2014-2024, respectively. Reports of tornadoes have increased by 2009, so as waterspouts by 2019. The highest tornadic activity was reported in 2022, with a total of events (37 tornadoes and 23 waterspouts), while the highest total activity is on the current year 2024 (36 tornadoes, 18 hail reports, 40 waterspouts). The low number of events reported in 2010 and 2016 is attributed to the initially applied methods used in the updated database and the low density of reliable information sources. The entire tornado and waterspout dataset shows a maximum of 3 tornadoes on the same day.

In recent years there has been an incremental trend in tornado activity records in US and European databases (Verbout et al., 2006; Groenemeijer and Kühne, 2014; León-Cruz et al., 2022). This apparent increase in the activity could be associated with the addition of more official databases and the higher public interest in events with high social impacts (Antonescu et al., 2017). The same goes for the Philippines, it is inferred that such behavior is associated with enhanced social attention to these natural phenomena, the extensive use of social media networks, and increased access to internet services.

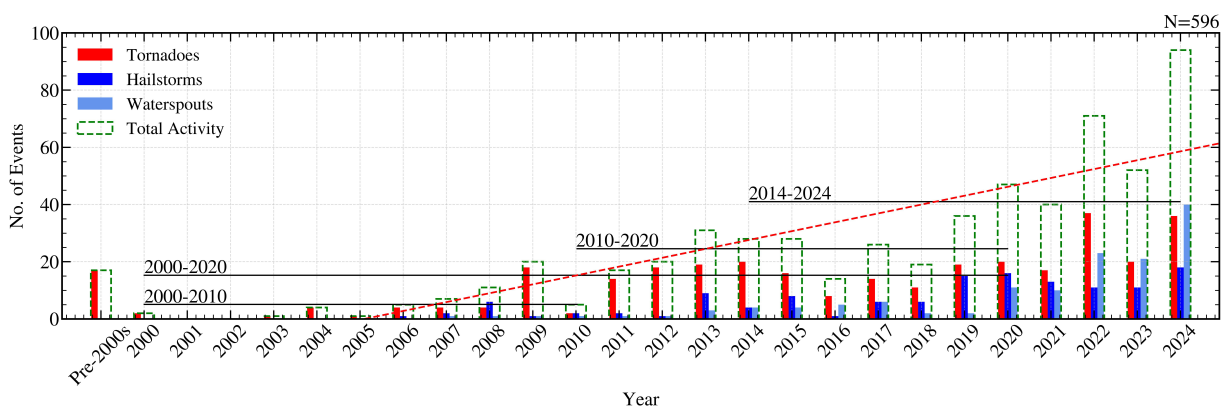

**Figure 2.** Annual distribution of documented tornadoes, hail, waterspouts, and total activity in Philippines for the period 1969-2024. The black solid lines indicate annual averaged activity for different periods. The red dashed line corresponds the linear $R^2$ trend from 2004-2024 period.

The highest severe weather activity is recorded in the warm spring and summer period, which starts between March-April with break-even counts at first, then severe weather season underway (Fig. 3a). The prolonged severe weather season, which includes tornadoes, hail events, and waterspouts, extends from May to August (Fig. 3a) winding down by September. During summer (June, July, and August, JJA), the number of documented tornadoes oscillates around 50 cases mo$^{-1}$ (Fig. 3a), while hail events decline throughout the said period. In this season, the arrival of tropical waves and onset of southwesterly winds to Philippines (Cruz et al., 2013; Dominguez et al., 2021) favors the moisture transport from oceans to the continental areas. The quality of low-level moisture leading to instability and ambient wind profile i.e. wind shear are the primary ingredients for robust convective storms to initiate capable of producing severe weather hazards at such period (Rasmussen and Blanchard, 1998; Thompson et al., 2003; Brooks et al., 2003; Brooks, 2009; Davies-Jones, 2015; Prein and Holland, 2018; Tang et al., 2019; Nixon et al., 2023).

Furthermore, the maximum registers of waterspouts are also on JJA period with August being at the highest (Fig. 3a). A lag period is evident when comparing the occurrence of the typical tornadoes and waterspouts. Interestingly, the active season of waterspouts coincides with the seasonal occurrence of tropical cyclones (TCs, Teng et al., 2021; León-Cruz et al., 2022). In this sense, a relationship can be established between the TC and the waterspout seasons. Although such an association is known in other parts of the world (Weiss, 1987), it has not previously been recognized in the Philippine context. Further research is needed to understand in-depth the implications of TC activities and waterspouts in the country and generally, the timing of these SWEs to other pre-existing weather systems.

Some tornadoes and waterspouts were documented in the cold winter period of December, January, and February (DJF) and in autumn (September, October, and November, SON). These events have been classified as cold tornadoes (Taszarek and Kolendowicz, 2013) and could be associated with a different dynamics than the conditions of the asian summer monsoon period. Seasonally, the number of documented SWEs between November and February reduces (an average of 11 SWEs yr$^{-1}$) with average of 8 tornadoes yr$^{-1}$, 4 hail events yr$^{-1}$, and 2 waterspout yr$^{-1}$ on that time span and can be related to stability co-nditions derived from the cold air masses from the Northeast Monsoon reducing moisture fluxes across the archipelago, mainly in Luzon landmass.

In summary, the severe weather season kicks off in spring (March, April, and May, MAM), reaches its peak in summer (JJA) along with reported hail events decreasing from its peak in the halfway of those seasons (AMJ; Selga, 1929), accompanied by the winding down of SWEs in autumn (SON), and finishes in winter (DJF) (Fig. 3b). Between MAM and JJA, on average, around 225 SWEs were reported encompassed by more than 100 tornadic activities and in the range of 50 events for both waterspout and hail activity within the parts of Visayas and Luzon (to be discussed). There is an evident difference between tornadoes, commonly reported in spring and summer, and waterspouts, which have their active period between summer and autumn. It is inferred that waterspouts are notably influenced by tropical cyclone activity, principally along the eastern coast of the Pacific Ocean. On the other hand, tornadoes are more related to instability conditions derived from the pass of easterly waves, moisture advection, favorable wind profile, and the shear-line activity[3].

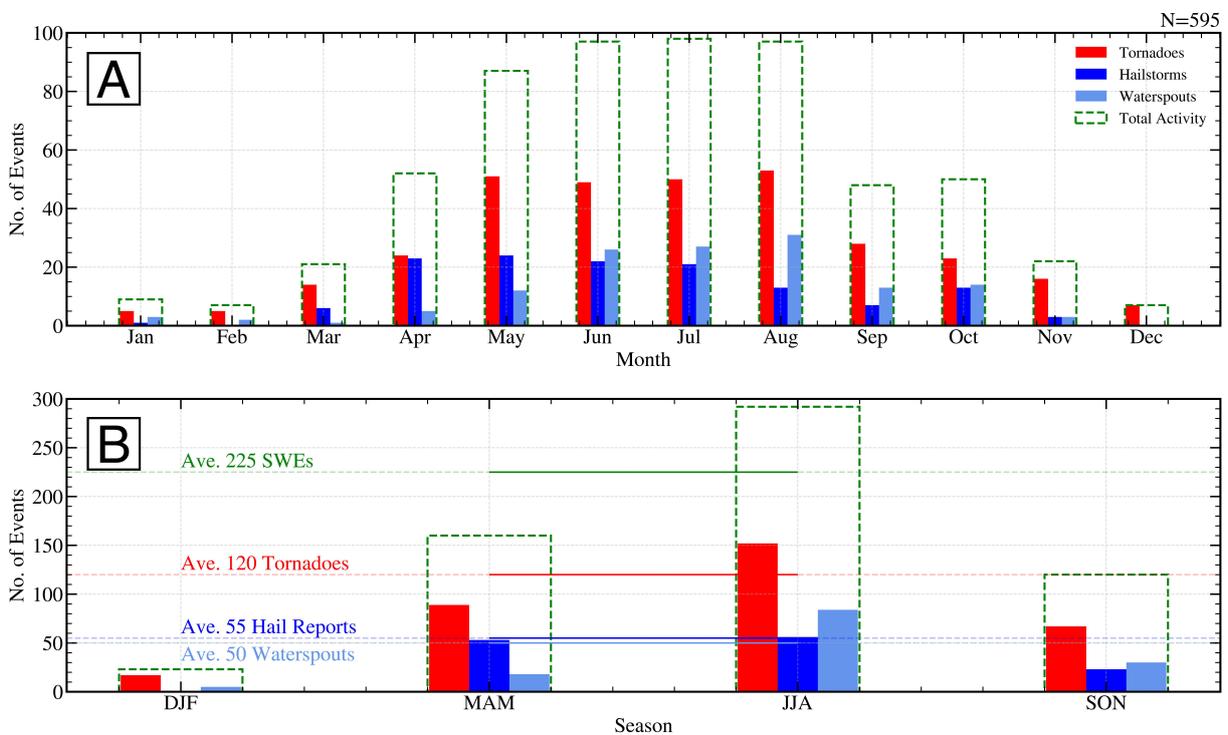

**Figure 3.** (a) Monthly and (b) Seasonal distribution of documented tornadoes, hail, waterspouts, and total severe weather activity in Philippines. In Fig. 3b, solid color lines represent the averages between MAM and JJA periods.

The diurnal distribution (Fig. 4) shows that these phenomena are usually reported between 14:00 and 15:59 h local time (06:00-07:59 h UTC), followed by 16:00-17:59 h local time (08:00-09:59 h UTC). This behavior is related with daytime heating and development of air-mass thunderstorms (Dai, 2001), necessary for convection to persist, hailstones to develop along the updrafts, and tornadogenesis. The diurnal distribution of severe weather activity reported in Philippines matches the findings in several climatologies (Groenemeijer and Kühne, 2014; Chen et al., 2018; Miglietta and Matsangouras, 2018; León-Cruz et al., 2022) and rainfall climatologies by Banares et al. (2021).

---

3 Also known as Tail-end of a Frontal System, usually a Cold Front.

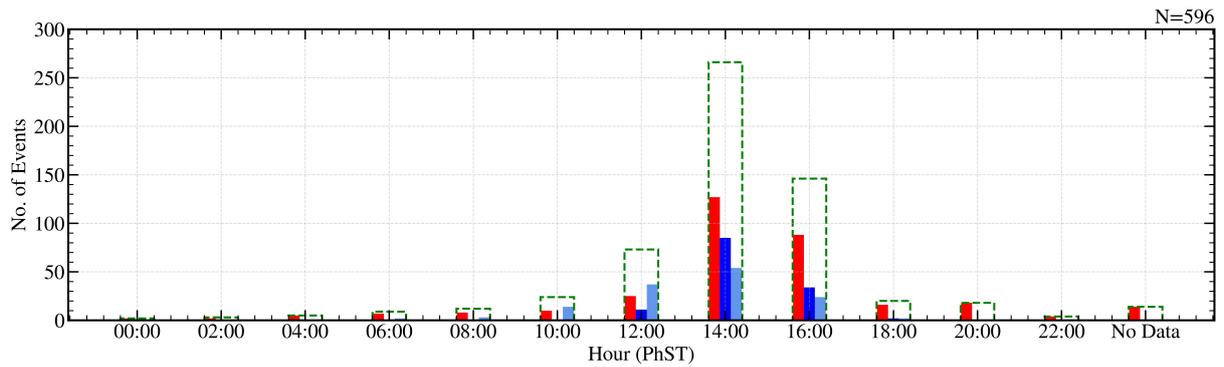

**Figure 4.** Diurnal distribution (local time) of archived tornadoes, hail, waterspouts, and total severe weather activity in the Philippines.

### 3.3 Spatial Distribution of Severe Weather Activities

The spatial distribution of SWEs is a crucial aspect of risk management associated with this phenomenon. Geographical characteristics and meteorological conditions are features to be considered in every baseline climatology for SWEs.

Figure 5a shows the geographic distribution of all documented SWEs across the archipelago. The highest tornadic and hail activity is registered in Greater Metro Manila (GMM) area encompassing Southern Luzon, National Capital Region (NCR), and Region III (Central Luzon). While scattered area of waterspout events were seen in along the coast of Southern Luzon and even at Laguna Lake, notably with the quadruple spouts back in May 2020 at height of pandemic. On the other hand, a large swath of waterspout events coinciding with few tornadic events was recorded in the Regions VI and VII (Western and Central Visayas). Lastly, another area where significant number of tornadoes were recorded was evident along BARMM (Bangsamoro Autonomous Region in Muslim Mindanao) and Region XII (SOCKSARGEN) with North and South Cotabato, Sultan Kudarat, and Maguindanao as its favored corridor for severe weather activity, making it $2^{nd}$ on the highest tornadic activity. Notably, a regional tornado outbreak occurred in South Cotabato at the afternoon of 30 September 2009 impacting several municipalities, including Koronadal City.

There is a clear association between population density and documented tornadoes (Figs. 5a and 5b). More than 50% of the Philippine population are located where the SWEs are recorded. The Region IV-A, NCR, and Region III takes up 30%[4] of the hazard providing a hint to the number of the documented SWEs (48%) were located in the Luzon landmass (Fig. 6). The population density effect on tornado and other forms of severe weather activity detection is well known (Anderson et al., 2007) and seen in this case. In addition to the population factor, the improvement in internet coverage and smartphone devices explains the increase in the number of reported tornadoes in recent years. This situation implies that the actual number of cases has been underestimated. Even though the aforementioned regions along Visayas down to Mindanao may have present topography and environmental characteristics similar to those observed throughout the GMM, the number of tornadoes, hails, and waterspouts documented is lower as seen in Fig. 6.

Figure 5c shows the climatological mean flash rate density observed by the Lightning Imaging Sensor (LIS) aboard Tropical Rainfall Measurement Mission satellite (TRRM, Kummerow et al., 1998; Cecil et al., 2014). Some studies have documented relationships among large hail, thunderstorms, and tornadoes e.g. Taszarek et al. (2020). The flash rate climatology presents a pattern most similar to SWE distribution across the GMM and Mindanao, however average flash rate density slightly plummets in the potential Visayas influence/hotspot zone. The convective processes are intensified over those regions, influenced by the upscale growth forming mesoscale convective systems (MCS; Lagare et al., 2023b)[5] and Southwest Monsoon (Cruz et al., 2013). These spatial variations could result from the differences in data sources. While the tornado and waterspout information come from reports by inhabitants, the flash data come from satellite-mounted lighting sensors.

---

4 Based on 2020 Census of Population and Housing by the Philippine Statistics Authority.

5 However, MCS were also reported along Mindanao's favorable area for tornadic storms. See Lagare et al. (2023a).

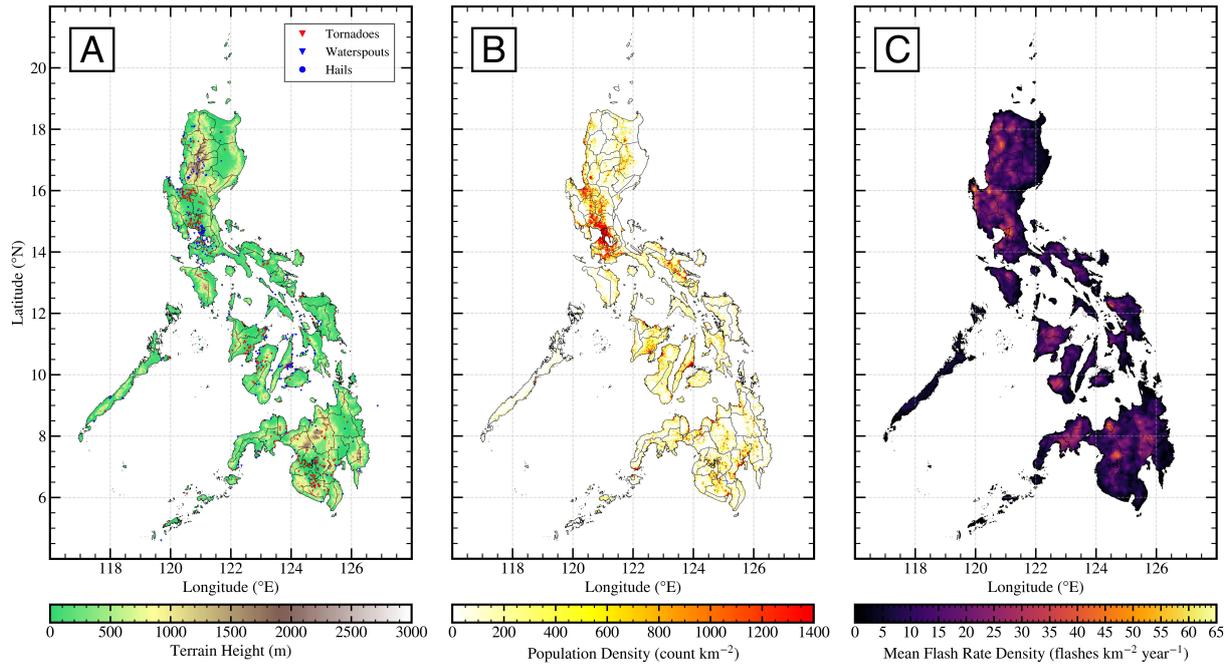

**Figure 5.** Geographical distribution of (a) documented tornadoes (red triangles), hails (blue circles) and waterspouts (blue triangles) in the Philippines from SWAP DR2, (b) Philippine population density from CIESIN (2020), and (c) climatological mean flash rate density from LIS-TRMM (1998-2014, Kummerow et al., 1998; Cecil et al., 2014).

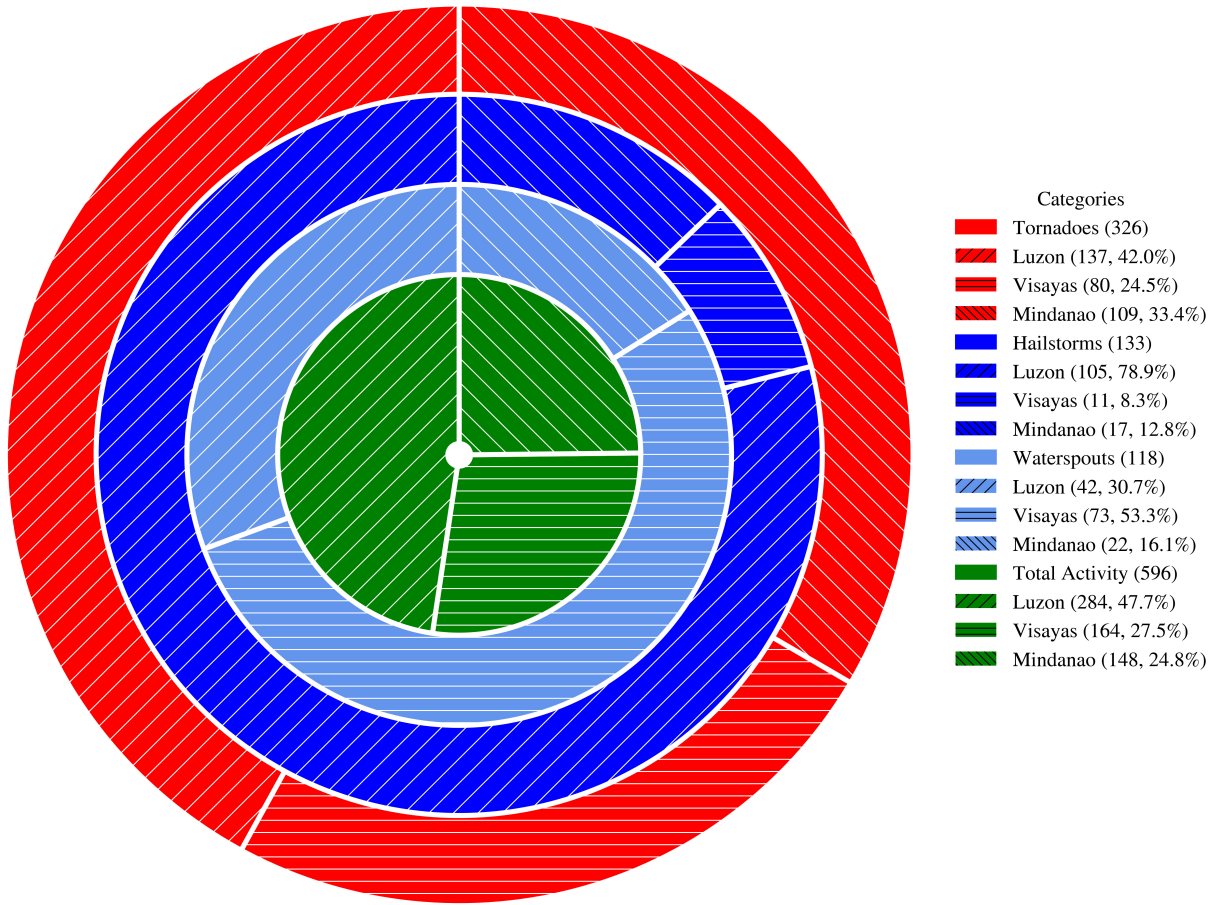

**Figure 6.** Percentage distribution of documented tornadoes (red), hail events (blue) and waterspouts (light blue), and total severe weather activity (green) across Luzon (//), Visayas (=), and Mindanao (\\) from SWAP DR2.

Figures 7a, 7b, and 7c displays a kernel density estimation made from *sci-kit learn* Python package on 11 km × 11 km grid. A clear and well-defined hotspot for all hazards threat (tornadoes, hails, and waterspouts) with bimodal distribution was latched in the GMM. The greatest likelihood for severe weather at the upper-portion of Metro Manila-Pampanga area and another at Pangasinan, while another hail hotspot was denoted along Benguet region (Figs. 6a and 6b). A total of 137 tornado and 105 hail reports have been documented over this area i.e. ≥ 40% and ≥ 70%, respectively of the total within SWAP DR2. Some of the notable events include the first recorded tornado in the Philippines back on 1968 June by Grazulis (2001), a potential EF2 tornado that tore parts of Bulakan, Bulacan on 1998 August, the well-documented EF1 Manila Tornado on 2016 August, the quadruplet waterspouts in Laguna Lake on 2020 May, a significant 8 cm hail in Norzagaray, Bulacan on 2021 August, and yet another Pampanga Tornadoes that impacted towns of Magalang and Arayat back on 2023 June and May of this year, respectively.

This specific hotspot in Luzon is surrounded by complex terrain features. In particular, to the west is form Zambales Mountain Range (ZMR) and to its east is the Sierra Madre Mountain Range (SMMR). Previous researches show that for landfalling tropical cyclones (LFTC), the complex terrain is crucial for magnifying rainfall across Luzon through orographic effects (Lagmay et al., 2015; Racoma et al., 2016). Although these were applied to TCs instead, such geographic features may also be conducive for building instability favorable for occurrence of severe weather in the area both on TC and non-TC days.[6]

In Visayas regions, two small tornado hotspots were discovered, overlayed by a strong signal of waterspout events across parts of Region VI and VII shown in Figs. 7a and 7c (as also depicted and discussed earlier). In particular, the Panay Islands, Negros Provinces, and much of Central Visayas shows a notable signature of SWEs focused along waterspouts. Two potential EF2 tornadoes were recorded in this area, both occurred in Cebu back on 2013 – one in Minglanilla and the other at Lapu-Lapu City. Meanwhile, more than 100 waterspout events were documented in this area as well, which is mostly rated with confidence as EF0s. A significant portion of the waterspout reports correspond to short duration (maximum 20 min.). However, with an extraordinary number of documented waterspout events in these aforementioned areas of Visayas, compared to other severe weather hotspots in the archipelago, an extensive analysis is required to understand the environment surrounded by mostly bodies of water suitable for tornadogenesis and non-tornadogenesis.[7]

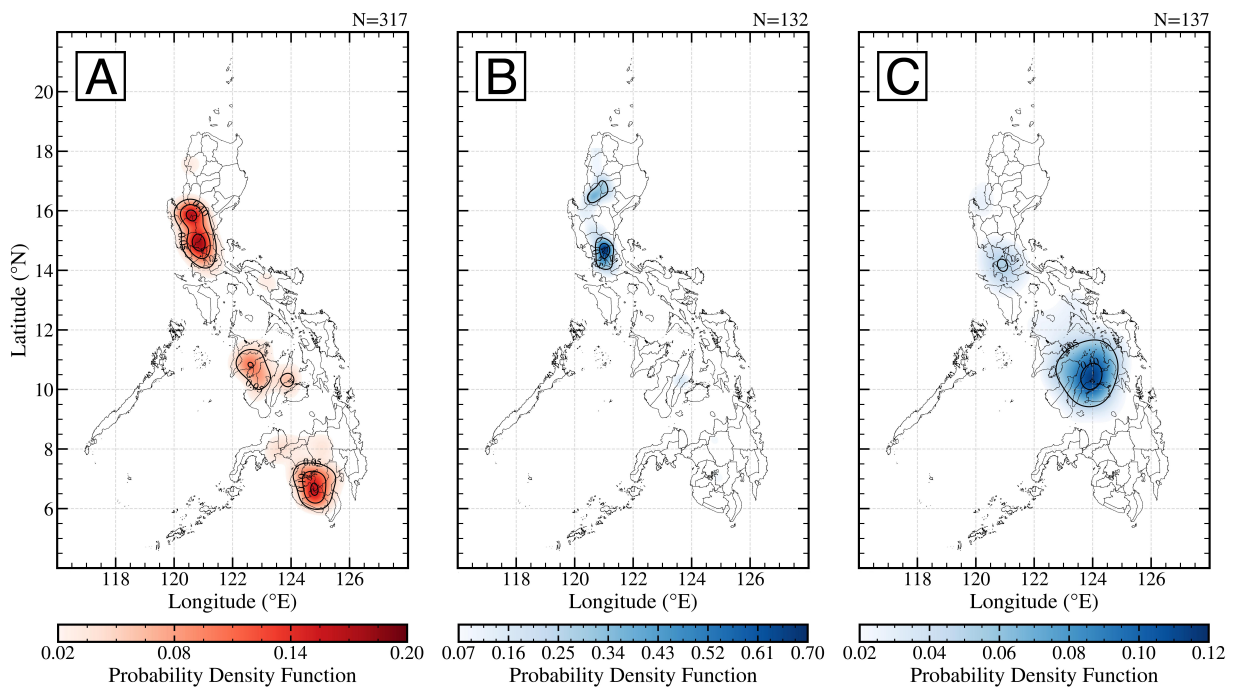

**Figure 7.** Kernel density estimate on a 11 km × 11 km grid of (a) tornado events, (b) hailstorm events, and (c) waterspout events from SWAP DR2. Labeled contour isoline are included at each panel. Heavier colors represent a higher SWE estimate.

---

6 This is most likely the case during SWM periods, as southwesterly (veering) winds blow over the ZMR. Due to conservation of potential vorticity, a lee-side cyclogenesis and its associated meso-low occurs at the backside of ZMR and seems to be the 'spark' for the Manila Tornado case as studied by Capuli (2024) and so as to other SWM period/JJA/warm summer season tornadoes along the hotspot zone in Luzon. On the other hand, prevailing easterly winds conditions during MAM season and monsoon breaks that can potentially lead to severe convective storms remains understudied.

In Mindanao, the data on tornado hotspots highlight a significant area of increased tornadic activity ($\geq$ 100 tornadoes), particularly in South Cotabato extending towards North Cotabato, with a weaker influence noted in Northern Mindanao (Figs. 7a and 7c). This region stands out due to its frequent occurrence of tornadic storms, contributing to its classification as a tornado hotspot. Interestingly, this increased tornado activity is situated between two major mountain ranges: the Pantaron Mountain Range to the east and the Highlands of Tiruray to the west. This geographical positioning may play a critical role in the development and intensification of tornadic storms in the area. The mountains could influence local wind patterns, moisture distribution, and the stability of the atmosphere, creating favorable conditions for tornado formation. This interplay between topography and weather patterns may partially explain why this region experiences higher-than-average tornadic activity compared to other areas in Mindanao.

Additionally, the region has a historical record of significant tornado events, including three potential EF2 tornadoes. Notably, a killer tornado struck Zamboanga del Sur in 1990 June, resulting in the deaths of up to 30 people in Manukan, despite being outside the main hotspot zone. Other destructive tornadoes were recorded in Lantapan, Bukidnon on 1991 July, a regional tornado outbreak in South Cotabato back on 2009 September, and in Pikit, North Cotabato in 2015 August. The severity and impact of these tornadoes further underscore the vulnerability of the region to SWEs, especially within the identified hotspot zone. However further studies could explore the precise meteorological mechanisms at play, enhancing the understanding of tornadogenesis in this complex terrain.[8]

Figure 8 shows the monthly spatial distribution of all tornadoes, hails, and waterspouts across the archipelago within SWAP DR2. The winter period (DJF) shows few and scattered cases of tornadoes across the entire archipelago. At the beginning of the active season in March, the number of cases increases with hail activity picking up in the Northern Luzon, and tornado activities within Panay Islands, Maguindanao, and South Cotabato. By April, a dominant hail event pattern is observed across the Luzon Landmass, while tornado season had a head-start in the Mindanao influence zone. In May, there are relatively high concentrations of tornadoes and hail events along the bimodal hotspots in Luzon, mixed batch of severe weather along Regions VI and VII, and scattered severe weather activity within Mindanao as well.

At the initial peak on June, severe weather activity is already spread out across the country, with tornado and hail reports centered in Pangasinan and across GMM area. Also, waterspout and tornado season is underway in Western and Central Visayas region, so as in Mindanao. This pattern is maintained throughout July and August across the country, with increased waterspout activity along the Visayas regions. At September, severe weather season in Luzon is about to end with decrease of severe weather activity (decrease in tornado cases) and shifts down south, with notable clusters around Visayas and Mindanao hotspots. October registered some hailstorm activity across NCR and Baguio due to monsoon breaks shifting the wind pattern to easterlies and the initial arrival of Northeast Monsoon. Meanwhile, Visayas and Mindanao severe weather season is still on going along Negros Provinces and Cebu, and along the Mindanao tornado hotspot as depicted earlier. By November, severe weather activity caps off in Luzon as Northeast Monsoon starts to advance and effect the landmass, so is about to end for Visayas with fewer registered SWEs, however Mindanao tornado activity remains quite active, surprisingly. By December, the severe weather activity flattens across the archipelago with few documented cases.

# 4. Conclusions

This paper has presented the first baseline climatology of SWEs encompassing tornadoes, hails, and waterspouts in the Philippines. Project SWAP and its DR2 consists of previous literature and recent documentary datasets, with a temporal coverage of 56 yr (1968-2024). The archive has many potential uses: its global reach allows for worldwide estimates of severe weather climatology, even the intercomparisons of severe weather observation and documentation methodologies. We hope that the digitized dataset will open possibilities for more broad climatological studies. More importantly, this project and paper itself can serve as a foundational piece for aspiring Filipino researchers trying to get a grasp and would like to study meteorology, and SWEs such as tornadoes within Philippine context in the future. Still, any analyses using this archive will require careful consideration of biases therein, many of which we have discussed.

---

7 Most of these Visayas-located waterspouts are rotating anti-cyclonically while few cases rotate cyclonically, based on documented videos linked in SWAP DR2.

8 As also noted the same for the severe weather hotspot in Luzon.

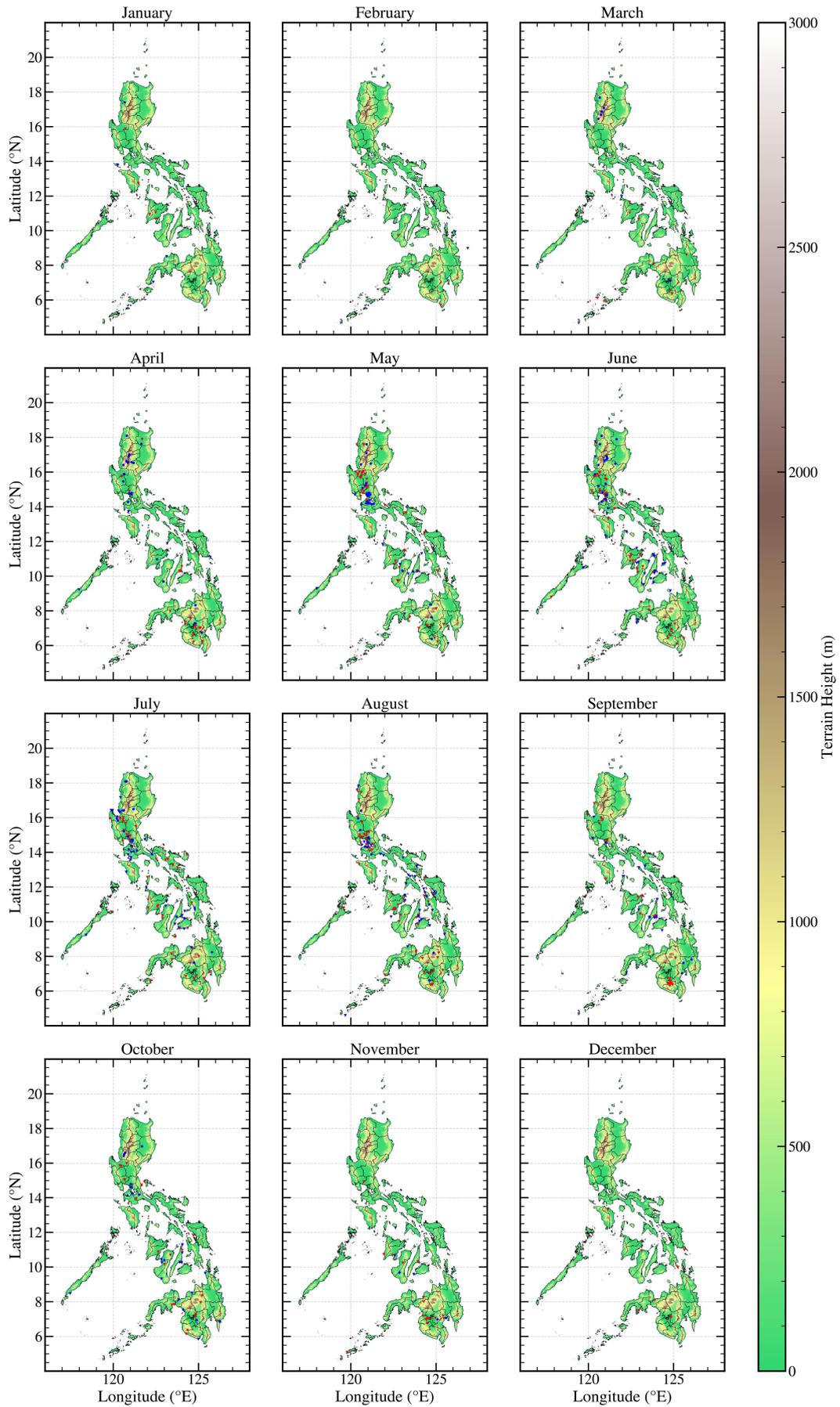

**Figure 8.** Monthly geographical distribution of documented tornadoes (red triangles), hail events (blue circles) and water spouts (blue triangles) in the Philippines from SWAP DR2.

On the analysis side, the increase in the number of documented SWEs is attributed to the rise in public awareness of these natural phenomena, the growth in the use of social networks (e.g. Facebook, Twitter, and YouTube), and improvements in information technologies (e.g. internet access). Such an apparent increase in tornadic, hail, and waterspout activity in Philippines could be associated with natural variability as well. For example, previous studies have been showing the influence of the El Niño-Southern Oscillation (ENSO) and the Madden-Julian Oscillation (MJO) in convection processes and precipitation (Lee et al., 2016; Molina et al., 2018; Tippett, 2018). However, it is not yet possible to determine the influence of such oscillations on tornado formation in Philippines, as there is insufficient evidence and relatively short data series.

The monthly distribution shows that the beginning of the most active phase of the severe weather season is in April. This first activity fits with the most active tornado phase in the USA (Taszarek et al., 2020). In May, SWEs have been previously documented across the country, in particular at lower Region I and Region III encompassing GMM area and Southern Luzon, with early termination between September-October. This may be indicating that this portion of the country may be related to similar atmospheric processes that lead to supercell formation and tornadogenesis, especially during Southwest Monsoon periods and the start of TC season in the western Pacific. However, easterly wind setups during months of April and May may also provide the same hazardous meteorological condition. Several studies documented the characteristics and role of tropical activity in rainfall in Philippines (Cruz et al., 2013; Racoma et al., 2016; Bañares et al., 2021). However, the impact of TCs on tornadogenesis and severe weather initiation remains understudied in the Philippine context.

Meanwhile, down south towards the Western and Central Visayas, waterspouts, along some tornadoes in the Panay Islands, are the main show for severe weather activity. The start of increased waterspout and tornado activity is in May followed by scattered and clumps of activity across the island groups of Visayas until it slowly winds down by November. On the other hand, Mindanao, particularly BARRM and Region XII, had a head start in terms of severe weather; mostly tornadic events, from April to November with severe weather season capping off by December across the country. The environments in these locations were generally comprised in accordance to the Philippine Climate Types developed by Coronas (1920) and Kintanar (1984), but the convective and kinematic setup of these remains to be studied.

Project SWAP and its DR2 depicts that nearly 50% of cases are located in the Pangasinan-GMM influence area. This fact may involve 2 critical issues: First, there is a clear relationship with population density. Demographic effects on tornado documentation have been previously reported (Anderson et al., 2007; Dias, 2011; Groenemeijer and Kühne, 2014; León-Cruz et al., 2022), and Philippines is subject to such trend. Second, this spatial pattern could also be related to topographic features. Much of these hotspots, pertaining to the Luzon and even Mindanao hotspots, were located in between two mountainous areas. Topographic effects were also reported along Luzon (Lagmay et al., 2015; Racoma et al., 2016), but seldom if not none, along Mindanao hotspot zone. However, a previous study on an MCS (Lagare et al., 2023a) may give an insight to the role of complex terrain along the said hotspot and for the increased tornadic activity in the area of interest.

The overall results and conclusion in this pioneering paper highlight that a baseline climatology for SWEs also provide a crucial reference point for tracking future changes in storm dynamics due to anthropogenic climate change (ACC). In general, ACC is expected to increase the Convective Available Potential Energy (CAPE) by increasing temperature and humidity within the atmospheric boundary layer (ABL) while simultaneously weakening vertical wind shear (typically, the 0-6 km Shear) by decreasing the pole-to-equator temperature gradient (by following the thermal wind equation/relation; Trapp et al., 2007)[9]. Recall that these two parameters typically characterize the meteorological conditions that foster severe thunderstorm formation and their entailed hazards. In fact, this was supported by a recent climatological study using reanalysis by Taszarek et al. (2021) who concluded that robust decreases are observed in the 0-6 km Bulk Shear variable across Southeast Asian regions, while CAPE trend increases in this region.

---

9 The thermal wind relation is expressed as; $\frac{\partial \vec{V}}{\partial Z} \cong \frac{g}{fT} \hat{k} \times \nabla T$ where $f$ is the Coriolis parameter and $g$ is the gravitational accelera-
tion, owing to projected weakening of the horizontal gradient of temperature ($\nabla T$).

However, when jointly evaluated, the increase in CAPE more than compensates for the decrease in shear such that the environment would still be considered favorable for severe convection capable of producing all-hazards threats. General circulation model (GCM) and regional climate model simulations reveal decreases in VWS that are disproportionately smaller than increases in CAPE, indicating an increase in frequency and/or intensity of future SWEs under ACC (e.g. Trapp et al., 2007; Woods et al., 2017; Raupach et al., 2021; and references therein). Upon checking some of the sounding profiles of these archived SWEs, their 0-6 km shear magnitude generally plays $\sim 10 \text{ m s}^{-1}$ while having sufficient instability (CAPE > 2000 J kg$^{-1}$)[10]. A complete, thorough/rigorous mesoscale analysis of convective and wind profile setting of these influence zones we identified in the Philippine context has been proposed.

What is more striking is that the locations of these severe weather hotspots are well-placed to the climate types, specifically along Climate Type 1 and 3. As a quick discussion, Climate Type 1, so as Type 3 at one point, was characterized by pronounced wet season influenced by the southwesterlies from May to October and the dry season from November to April caused by the prevailing winds in form of north Pacific easterlies, with maximum rainfall during June to September (Jamandre and Narisma, 2013; Peralta et al., 2020). However, the timing of the wet season can vary depending on the onset of the southwesterly wind that flows at the country, tied to the Asian summer monsoon. Still, aiding on the initiation convective activity over the western coast of the Philippines during aforementioned period (Akasaka, 2010).

Now, this begs the question; what are the convective and kinematic mechanisms within these climate types, not based on rainfall and temperature, making it favorable environment for severe weather? *What makes the clock tick* capable of producing these SWEs? As proposed and a Part 2 of this project, we will explore and establish another baseline climatology, this time centered around hazardous convective weather setups across the archipelago. Given that Project SWAP and its DR2 was recently distributed, SWEs continue to appear and consequently be listed in the archive. No doubt, that a 3$^{rd}$ Data Release (DR3) will be coming soon with refinement to the data archiving and integration flow. As SWEs also continue to rise within the archipelago (as of November 2024, more than 1000 SWEs are already tallied), in the near future, an update to the climatology of SWEs across the archipelago is also proposed and will be part of the continuation of this paper.

Lastly, the presence of Project SWAP does not take away in any way the critical role served by DOST-PAGASA offices scattered across the country. Instead, like any other archiving projects and their ensuing results, such an agency should provide expert severe weather forecasting services to create a critically relevant "safety net" of guidance products in support of the national and regional forecasting services throughout Philippines. Given that such an agency would have only to concern itself with the specific severe weather events under its charter, it would serve essentially to support all the national and regional forecast services, whose attention is often dispersed among a wide diversity of weather forecasting responsibilities (e.g. aviation weather, agricultural forecasts, routine public forecasts, interacting with the public, etc.). Although frequencies of these SWEs are lower than the U.S. Great Plains and European continent (e.g. Smith et al., 2012; Groenemeijer and Kühne, 2014; Allen et al., 2019; and references therein), another obvious conclusion to this study is that severe thunderstorms; whether supercellular or non-supercellular, and their entailed hazards are common in Philippines than most Filipinos realize. As proposed above, understanding their environments is a top priority.

Thus, we want to emphasize that the importance and opportunity to create a new infrastructure to meet uniquely requirements represents an important challenge to the Philippine meteorological community given this recent results regarding severe weather climatology. By being aware of the American example such as the NWS' Storm Prediction Center (SPC) and their previous-to-current suite of products for severe weather forecasting (e.g. Rapid Refresh/Rapid Update Cycle: RAP/RUC models, Early Warning Systems, Hailpads, Tornado Shelters etc. even severe weather infrastructure in other countries), it is possible to avoid at least some of the mistakes made during the largely ad hoc development of the American system. Galway (1989) pointed out that the system begun when political demand for it arose, and then ad hoc solutions, put together in order to respond to that political pressure, eventually solidified into the existing system. If it were possible to create a new meteorological infrastructure for dealing with tornadoes, hailstorms, and waterspout events and their insuing hazards on the basis of a systematic study, the result might differ in important ways from the existing structure in the Philippines rectifying efforts to mitigate the hazards' potential damages, fatilities, and economic losses.

---

10 Example of such mesoscale setup is the case of August 13, 2021 Hailstorm in Norzagaray, Bulacan (Capuli, *in preparation*).

To do such, the public and private sector forecasters, news media and their broadcasting meteorologists, governmental emergency management agencies such as in the Disaster Risk Reduction Management (DRRM) sector, public and private agencies for dealing with post-disaster situations, the politicians (who must provide the resource support to make services available), and even research scientists of all sorts, as well as structural engineers must be "partners", not the other way around i.e. competitors or enemies to be effective in dealing with severe weather hazards. Creating and updating an effective severe weather infrastructure to cope with SWEs depend critically on forging and maintaining good working relationships among all these components of an effective system.

Project SWAP and the community project leader (CGHA, corresponding author) hopes that future decisions about the aforementioned infrastructure will not be planned and made hastily, but rather with due care as the impacts of weather on society are complex and do not inevitably have equal impacts from case to case.

**Data Availability Statement.** Project SWAP and its SWAP DR2 is available and distributed on Zenodo. The Digital Elevation Model (DEM) is from SRTM15Plus distributed and available on OpenTopography. The population density is available through World Population Hub. Finally, the lightning data from the TRMM is accessible through NASA Earth Data. All of these data are distributed under Creative Commons CC-BY 4.0. Proper attribution is required for these datasets. This paper has made of use of the following Python packages: *sci-kit learn*, *Matplotlib*, *GeoPandas*, *Pandas*, *Rasterio*, *rioxarray*, *NumPy*, and *Regionmask*.

**Acknowledgements.** We are very grateful to the Philippine population for reporting the occurrence of tornadoes in their communities. We also appreciate the valuable comments of the three anonymous reviewers and editor, which helped to improve this manuscript. This work received no funding. However, it was made possible through extensive and exhaustive effort, whose dedication and commitment to advancing our understanding of severe weather phenomena were indispensable. We would also like to extend our appreciation to the government agencies, acting as the primary source, for providing essential data that contributed significantly to this work. Finally, we are thankful to our families and loved one for their unwavering support throughout the project.

*CORRESPONDING AUTHOR: Generich H. CAPULI,
Rizal Technological University, Brgy. Malamig, Mandaluyong City, Metro Manila 1550, Philippines
e-mail: genhcapuli@rtu.edu.ph